# Heteroepitaxial growth and optoelectronic properties of layered iron oxyarsenide, LaFeAsO

Hidenori Hiramatsu [a, *], Takayoshi Katase [b], Toshio Kamiya [a, b], Masahiro Hirano [a, c], and Hideo Hosono [a, b, c]

(a): *ERATO–SORST, Japan Science and Technology Agency (JST), in Frontier Research Center, Tokyo Institute of Technology, S2-6F East, Mail-box S2-13, 4259 Nagatsuta-cho, Midori-ku, Yokohama 226-8503, Japan*

(b): *Materials and Structures Laboratory, Mail-box R3-1, Tokyo Institute of Technology, 4259 Nagatsuta-cho, Midori-ku, Yokohama 226-8503, Japan*

(c): *Frontier Research Center, S2-6F East, Mail-box S2-13, Tokyo Institute of Technology, 4259 Nagatsuta-cho, Midori-ku, Yokohama 226-8503, Japan*

PACS No.: 81.15.Fg (Laser deposition), 81.05.Zx (New materials: theory, design, and fabrication), 74.78.–w (Superconducting films and low-dimensional structures)

## ABSTRACT

Epitaxial thin films of LaFeAsO were fabricated on MgO (001) and mixed-perovskite (La, Sr)(Al, Ta)$O_3$ (001) single-crystal substrates by pulsed laser deposition using a Nd:YAG second harmonic source and a 10 at.% F-doped LaFeAsO disk target. Temperature dependences of the electrical resistivities showed no superconducting transition in the temperature range of 2 – 300 K, and were similar to those of undoped polycrystalline bulk samples. The transmittance spectrum exhibited a clear peak at ~0.2 eV, which is explained by ab-initio calculations.

Footnotes:

[*] Electronic mail: h-hirama@lucid.msl.titech.ac.jp

Since the recent discoveries of superconduction in F-doped LaFeAsO [LaFeAs(O, F)], [1,2] extensive studies on iron-based layered oxypnictides LnFePnO (Ln = lanthanides, Pn = pnicogens) have been performed, and the maximum transition temperature ($T_c$) has reached 55 K in SmFeAsO. [3] The finding of higher $T_c$ superconductors is expected in this and related material systems, [4] which provides an opportunity to realize macroscopic quantum coherence devices operating at practically high temperatures. For such devices, high-quality epitaxial films are inevitable because they require flat interfaces and stacking of thin layers at the atomic scale. High-quality epitaxial films are crucial also to know intrinsic properties of these new superconductor systems because the materials properties reported to date have been limited to bulk polycrystals except for a few works based on single-crystals. [5]

In this letter, we report epitaxial film growth of LaFeAsO and their electrical and optical properties. In the course of this work, we have found that, unlikely other isostructural compounds of LaFeAsO, it was difficult even to produce the LaFeAsO phase in deposited films. Therefore, we applied some modifications to a simple pulsed laser deposition (PLD) process.

All the film depositions were conducted by PLD using 10 at.% F-doped LaFeAsO targets to grow LaFeAs(O, F) films. First, we examined many processes including simple PLD growth, post-thermal annealing, and reactive solid-phase epitaxy (R-SPE), [6] which have successfully produced epitaxial films for isostructural compounds such as LnMXO (M = Cu or Zn; X = chalcogens or P), [7] BiCuXO, and BaCuXF. [8] An ArF excimer laser was used in the processes. However, we could not obtain even a film containing a small portion of the LaFeAsO phase. For example, films deposited by simple PLD at 600 °C were composites of impurity phases $La_2O_3$, $LaFeO_3$, and LaAs. Even at higher temperatures up to 800 °C, no LaFeAsO phase appeared and the wide bandgap impurities of $La_2O_3$ and LaOF increased as shown in a x-ray diffraction (XRD, Cu K$\alpha$ radiation) pattern [Fig. 1(a)]. We considered that higher temperatures, e.g. > 1000 °C, would be required to form the LaFeAsO phase similar to the synthesis of polycrystalline bulk samples. [1] Therefore, films formed by simple PLD at room temperature (RT) were subjected to post-thermal annealing at 900 – 1100 °C in evacuated silica glass ampoules, but the films consisted only of $LaFeO_3$, LaAs, and/or $La_2O_3$. When the R-SPE process was employed to crystallize amorphous LaFeAsO layers with an assistance of thin Fe metal layers as sacrificial template layers, although the thin Fe layer was reacted with the amorphous layer, only the impurities phases of $La_2O_3$, LaOF, LaAs, $Fe_2As$, and FeAs were produced. This difficulty in obtaining even the LaFeAsO phase is a sharp contrast to the cases of the other isostructural compounds mentioned above.

Finally, we succeeded in obtaining epitaxial films of LaFeAsO by a simple PLD, in which the excitation source for PLD was changed from the ArF excimer laser (wavelength: 193 nm) to the second harmonic ($2\omega$) Nd:YAG laser (532 nm). We speculated that the high energy excitation by the deep-ultraviolet ArF excimer laser produced high-density active oxygen species and thereby enhanced unfavorable oxidation of the metallic components. Besides, we would like to stress that another critical factor in this experiment would be high phase-purity of the PLD target. Small portions of impurities such as $La_2O_3$ and LaOF were detected in the targets used in the early stage described in the previous paragraph. The LaFeAsO phase has never been obtained in deposited films whenever these impurities were observed in the target: the impurities appeared to be selectively transferred to the film. After further optimizing the target, epitaxial films were obtained only when the target did not contain wide bandgap impurities such as $La_2O_3$ and LaOF. By contrast, inclusion of small amounts of narrow bandgap impurities such as LaAs and FeAs were not critical. Their concentrations were roughly estimated to be 1 mol% for LaAs and 9 mol% for FeAs by three-phase Rietveld analyses using powder XRD data. [9]

~300 nm thick LaFeAsO films were grown on MgO (001) single-crystal substrates using the $2\omega$ Nd:YAG laser (frequency: 10 Hz, fluence: ~1.5 J/cm$^2$) in a vacuum at ~$10^{-5}$ Pa. The substrate temperature ($T_s$) was varied from 700 to 880 $^o$C using a halogen lamp. Figure 1(a) shows the XRD patterns of the films as a function of $T_s$. When $T_s$ was 750 $^o$C, the LaFeAsO phase began to appear but its preferential orientation was very weak and it contained an impurity phase of LaAs, which is confirmed by the diffraction peak at $2\theta \sim 29^o$. At lower $T_s$, only a very weak and broad peak, which is difficult to be assigned to the LaFeAsO phase, was observed. With increasing $T_s$, the film preferential orientation became the strongest at 780 $^o$C, but the impurity LaAs content increased as well. The LaAs content monotonically increases and the preferential orientation became weaker with further increasing $T_s$, and another impurity phase of Fe metal started to be segregated at 880 °C. These results indicate that the optimum $T_s$ is ~780 °C. Although epitaxial films have been reproducibly obtained, the yield remains ~ 40 % at present, which would be due to the narrow process window and suggests there still are unknown control parameters to grow epitaxial films.

Since no superconducting transition was observed for the as-grown LaFeAsO epitaxial films, we also examined post-deposition thermal annealing in a 5% $H_2$ / Ar flow for 30 min in order to dope more electrons. We optimized the annealing condition using second-grade (i.e. weakly-oriented) LaFeAsO films that contain the same impurity phases found in the best film of Fig. 1(a) [Fig. 1(b)]. When the annealing temperature ($T_a$) was 400 °C, the LaAs impurity was removed, while other impurities such as $LaFeO_3$, FeAs and Fe appeared. For lower $T_a$, no change was observed in XRD patterns. With further increasing $T_a$, the peaks of the LaFeAsO phase broadened and the films were decomposed to an impurity phase at 700 °C (the peaks are shown by the vertical bars), where the decomposition product is not yet determined uniquely because diffractions of both $\gamma$-$Fe_2O_3$ and Fe-As-O compounds appear at similar angles. It was confirmed that 400 °C annealing in a pure Ar gas reproducibly removed the LaAs phase, and therefore the effect on removing the LaAs phase is attributed to the 400 °C annealing and not to the $H_2$ atmosphere. It is also important to note that the thermal annealing results are not explained by thermodynamics because $T_a$ was lower than $T_s$, and the remaining phases after annealing are largely different from those found in the as-grown films, suggesting that the formation of the LaFeAsO phase in the as-grown films at ≥ 750°C is assisted largely by nonequilibrium nature of the low temperature PLD process.

Figure 2(a) shows the out-of-plane high-resolution XRD (HR-XRD, Cu K$\alpha_1$ radiation) patterns of the LaFeAsO films. For the as-grown film on MgO, sharp 00$l$ diffraction peaks of LaFeAsO are observed along with MgO 002, LaAs $h$00, and cubic-Fe 110 diffractions, indicating the film is preferentially oriented along the $c$-axis, but contains these impurities. The full width at half maximum (FWHM) of the 003 rocking curve was 1.5 degrees. The in-plane XRD pattern [(i) in Fig. 2 (b)] of the as-grown film show LaFeAsO $hh$0 and MgO 220 diffractions only, and the LaFeAsO 110 diffraction shows a four-fold in-plane orientation in the in-plane rocking curve ($\phi$ scan), which corresponds to the tetragonal symmetry of the LaFeAsO lattice.[5] These observations substantiate that the film is grown heteroepitaxially on the MgO single-crystal and the epitaxial relationship is (001)[110] LaFeAsO ∥ (001)[110] MgO. The lattice parameters were $a$ = 0.4048 and $c$ = 0.8791 nm and larger (~0.3 % for $a$-axis, ~0.6 % for $c$-axis) than both of undoped and F-doped LaFeAsO bulks,[1] indicating that the expanded lattice parameters are due neither to the constraint of the substrate nor the F-doping. By $H_2$ annealing, the LaAs phase was drastically decreased by keeping the epitaxial relationship as shown in (ii) of Figs. 2(a, b). The lattice parameters were further increased by the $H_2$ annealing to $a$ = 0.4060 and $c$ = 0.8814 nm.

Next we fabricated LaFeAsO epitaxial films also on mixed-perovskite (La, Sr)(Al, Ta)$O_3$ (LSAT) (001) ($a$/2 = 0.387 nm) single-crystal substrates because MgO and LSAT substrates have different lattice mismatching with LaFeAsO (+4 % for the $a$-axis length of MgO and –4 % for the $a$/2 of LSAT). It was also expected that the use of LSAT would reduce the $a$-axis length of the film and assist appearance of a superconducting phase, because recent study demonstrated that the $T_c$ of LaFeAs(O, F) is increased from 26 to 43 K by decreasing the cell volume under a high pressure.[2] Similar to the MgO case, epitaxial films were obtained also on the LSAT with the same epitaxial relationship as seen in (iii) of Figs. 2(a, b). The FWHM of the 003 rocking curve was 0.8 degrees, which is better than that of the film on MgO. Oppose to our expectation, the lattice parameters ($a$ = 0.4053 and $c$ = 0.8791 nm) of the as-grown film on LSAT were similar to those of the film on MgO, indicating the atomic configuration at the substrate surface does not control the film growth. The films on LSAT are better than those on MgO also in the sense that the LaAs impurity was completely removed by the $H_2$ annealing [(iii) of Fig. 2(a)]. But the Fe metal remained even for the LSAT case. The lattice parameters were changed to $a$ = 0.4042 and $c$ = 0.8853 nm by the $H_2$ annealing. We also used MgAl$_2$O$_4$ (001) single-crystal substrates because the lattice parameter ($a$/2 = 0.404 nm) has the best matching with LaFeAsO among the single-crystals examined here. Similar out-of-plane and in-plane orientations were observed, but the diffraction peaks were weak, indicating crystallinity of the films were the worst irrespective of the very small lattice mismatch (0.1%).

Figure 3 shows the temperature dependences of electrical resistivities of the LaFeAsO epitaxial films along with those of polycrystals of undoped and 11 at.% F-doped LaFeAsO. [1] The resistivities of the as-grown films at 300 K are smaller than those of the polycrystals, suggesting that crystallinity of the epitaxial films is higher than those of polycrystals. The temperature dependence is nearly flat and tends to increase in the low temperature range, and the resistivities extrapolated to absolute zero have small values at the order of $10^{-2}$ Ωcm. By $H_2$ annealing, although the LaAs impurity was decreased and the electron density was expected to increase, the electrical resistivity increased. A superconducting transition was not observed in any sample grown in this study at least down to 2 K although polycrystals of LaFeAs(O, F) exhibit superconducting transitions at 10 – 26 K. [1] It is noteworthy that the temperature dependences of the LaFeAsO films look similar to those of undoped LaFeAsO bulks: i.e. small bumps appear at < 150 K, which corresponds to a structural transition from a tetragonal to orthorhombic lattice [10] accompanying a magnetic transition from paramagnetic (PM) to antiferromagnetic (AFM), [11] and the resistivities increase as temperature decrease, which is opposite to conventional metals. This similarity to the undoped samples suggests that F ions doped from the target do not work as donors or F ions may not be incorporated in the films.

Figure 4(a) shows a transmittance spectrum of a LaFeAsO epitaxial film on a LSAT at RT. Net transmittance ($T_{net}$) was evaluated from transmission ($T_{obs}$) and reflection ($R_{obs}$) spectra from 17000 to 4000 cm$^{-1}$ using an approximation $T_{net} = T_{obs} / (1-R_{obs})$ to correct the surface reflection loss. Note that metallic reflection due to Drude-type plasma oscillation is removed in $T_{net}$ and the $T_{net}$ reflects the pure absorption in the film. Infrared (IR) transmission was also measured with a FT-IR apparatus from 4000 to 1200 cm$^{-1}$. The net IR transmission spectrum was obtained by dividing the observed IR transmission spectrum by a transmission spectrum of the substrate. In the visible region at photon energies > 1.5 eV, $T_{net}$ was ~0 %, which is consistent with the black sample color. $T_{net}$ gradually increases with decreasing the photon energy, but even the maximum value is as small as 12 % and $T_{net}$ then decreases. Consequently, the $T_{net}$ spectrum exhibits a clear peak at ~0.2 eV. These features are very different from simple metals although the temperature dependence of resistivity is mostly metallic around RT. It is understandable from the electronic structures calculated based on the projector augmented wave method [12] and PBE96 functional using a code VASP. [13] These were calculated with optimized crystal structures by keeping the space group P4/*nmm*.

Figure 4(b) shows the total densities of states (TDOS) of two different spin ordering states, an AFM and a ferromagnetic (FM) phases. These calculations showed that the ground state is AFM and the spin moment of a Fe was 1.56 $\mu_B$ for the AFM phase and almost zero (< 0.05 $\mu_B$) for the FM phase, which are consistent with previously-reported results. [14] Therefore, the FM phase shown in Fig. 4(b) is essentially the Pauli PM phase and would be closer to the present case because the AFM transition temperature of undoped LaFeAsO is ~150 K. [11] Although the observed $T_{net}$ reflects the absorption coefficients in the *a-b* plane and is not quantitatively related to TDOS, this comparison shows that the $T_{net}$ peak at ~0.2 eV corresponds to the valley of the TDOS at 0 – 0.5 eV.

In summary, we succeeded in fabricating epitaxial films of LaFeAsO by PLD, and think that the employment of 2ω Nd:YAG and the wide gap impurity phase-free target were keys to forming the LaFeAsO phase in films and obtaining epitaxial films. A superconducting transition was not observed and the temperature dependence of resistivity was similar to that of undoped polycrystalline bulk samples, suggesting electrons were not successfully doped in the films. The epitaxial films revealed that the optical spectrum shows a clear peak at ~0.2 eV at RT, which is consistent with the result of ab-initio calculations.

We would like to thank Drs. K. Kawamura and Y. Kamihara of ERATO-SORST, JST and Prof. K. Kajihara of Tokyo Metropolitan University for their kind helps.

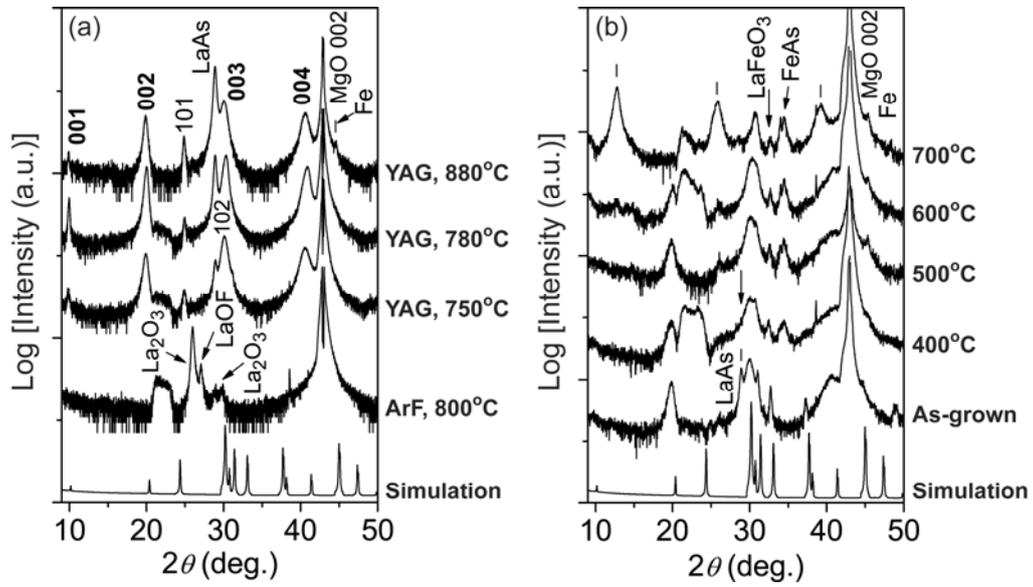

FIG. 1. XRD patterns of (a) as-grown films prepared using ArF excimer or Nd:YAG laser as an excitation sources of PLD, and (b) films annealed in a $H_2$ flow. Substrate and annealing temperatures are shown at right of each figure. The simulated powder pattern of undoped LaFeAsO is shown at bottoms for comparison.

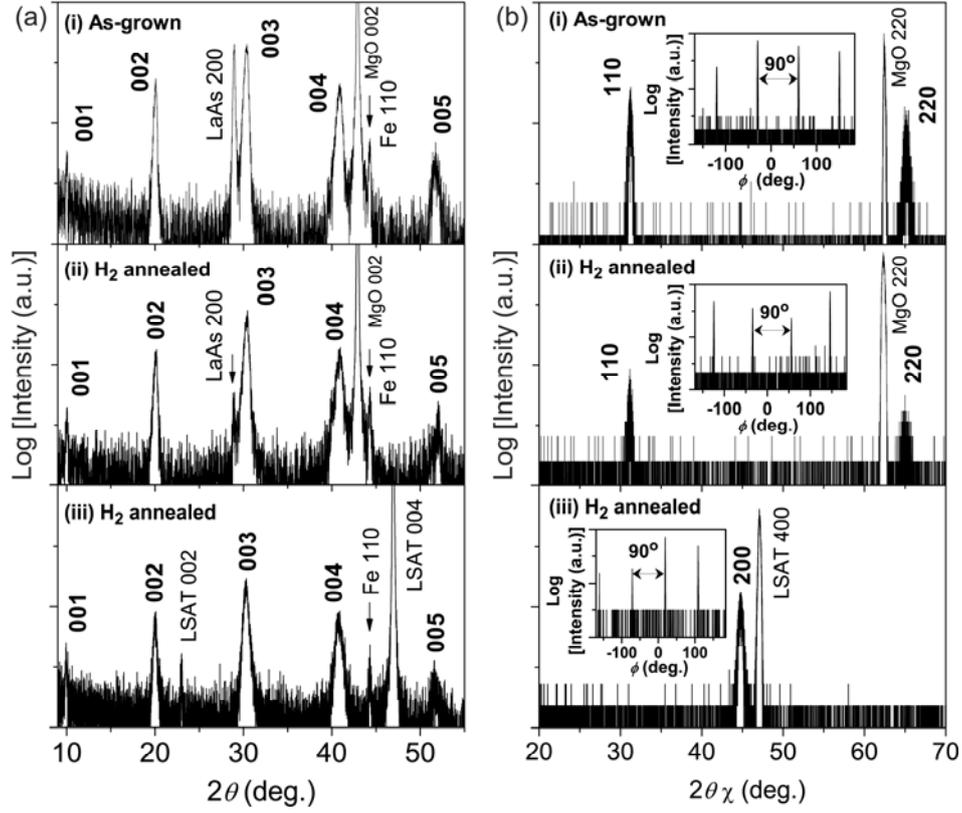

FIG. 2. HR-XRD patterns of LaFeAsO film. (a) Out-of-plane and (b) in-plane patterns. (i) As-grown film on MgO (001), (ii) $H_2$ annealed film on MgO (001), and (iii) $H_2$ annealed film on LSAT (001). Insets in (b) show the in-plane rocking curves ($\phi$ scans) of the 110 diffractions for the films on MgO and of the 200 diffraction for the film on LSAT.

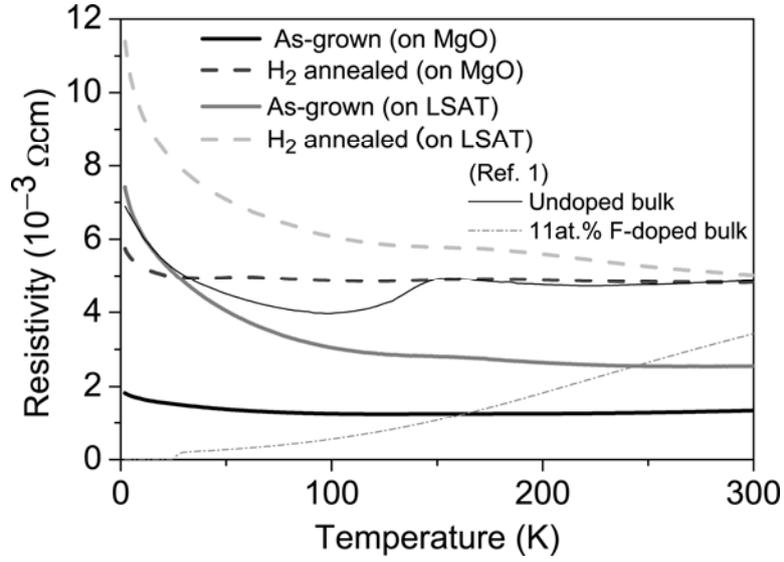

FIG. 3. Temperature dependences of electrical resistivities of LaFeAsO epitaxial films. Those of undoped and 11at.% F-doped polycrystalline bulks are shown for comparison.

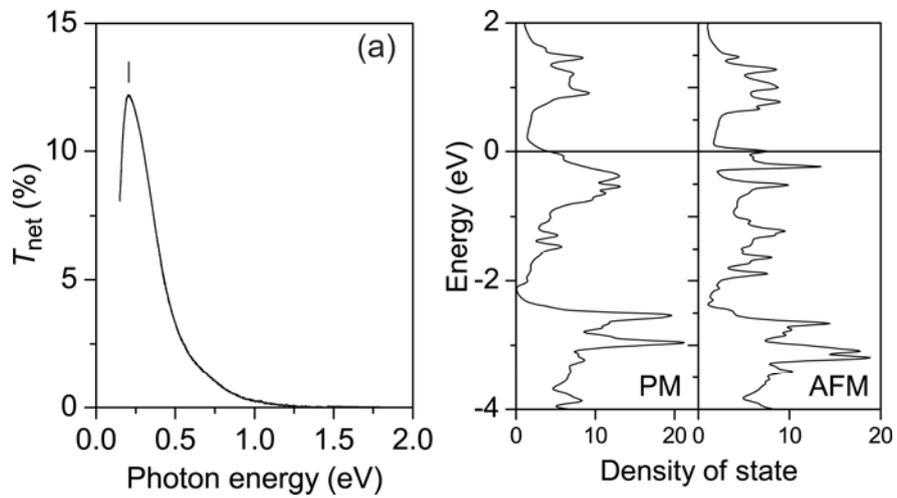

FIG. 4. (a) Optical transmittance spectrum of LaFeAsO epitaxial film at room temperature. (b) Calculated total densities of states for PM (left) and for AFM (right) phases. The energy is measured from the Fermi energy.